**Where is Memory Information Stored in the Brain?**

James Tee,[1,2] and Desmond P. Taylor[2]

[1]Department of Psychology, The New School for Social Research, USA.

[2]Department of Electrical and Computer Engineering, University of Canterbury, New Zealand.

james.tee@newschool.edu

WHERE IS MEMORY INFORMATION STORED IN THE BRAIN?

**Abstract**

Within the scientific research community, memory information in the brain is commonly believed to be stored in the synapse – a hypothesis famously attributed to psychologist Donald Hebb. However, there is a growing minority who postulate that memory is stored inside the neuron at the molecular (RNA or DNA) level – an alternative postulation known as the cell-intrinsic hypothesis, coined by psychologist Randy Gallistel. In this paper, we review a selection of key experimental evidence from both sides of the argument. We begin with Eric Kandel's studies on sea slugs, which provided the first evidence in support of the synaptic hypothesis. Next, we touch on experiments in mice by John O'Keefe (declarative memory and the hippocampus) and Joseph LeDoux (procedural fear memory and the amygdala). Then, we introduce the synapse as the basic building block of today's artificial intelligence neural networks. After that, we describe David Glanzman's study on dissociating memory storage and synaptic change in sea slugs, and Susumu Tonegawa's experiment on reactivating retrograde amnesia in mice using laser. From there, we highlight Germund Hesslow's experiment on conditioned pauses in ferrets, and Beatrice Gelber's experiment on conditioning in single-celled organisms without synapses (*Paramecium aurelia*). This is followed by a description of David Glanzman's experiment on transplanting memory between sea slugs using RNA. Finally, we provide an overview of Brian Dias and Kerry Ressler's experiment on DNA transfer of fear in mice from parents to offspring. We conclude with some potential implications for the wider field of psychology.

**Where is Memory Information Stored in the Brain?**

When we memorize a phone number, where is this information stored? Regardless of the type of memory (e.g., episodic, semantic, autobiographical) or the theories of memory (e.g., storehouse, reconstructive), memory information must be stored somewhere in the brain. Psychologist Karl Lashley's lifelong search for the answer ended somewhat fruitlessly: "I sometimes feel, in reviewing the evidence on the localization of the memory trace, that the necessary conclusion is that learning just is not possible" (Lashley, 1950, pp. 477-478). His work was continued by one of his PhD students, Donald O. Hebb, well-known today for his synaptic hypothesis:

> When an axon of cell A is near enough to excite cell B and repeatedly or persistently takes part in firing it, some growth process or metabolic change takes place in one or both cells such that A's efficiency, as one of the cells firing B, is increased. (Hebb, 1949, p. 62)

This hypothesis is often summarized as "cells that fire together wire together" (Shatz, 1992). The key idea is that changes in synaptic strength and connectivity may serve as the fundamental mechanism for information storage in the brain (Trettenbrein, 2016); that is, memory information is stored in the synapse. Some researchers, however, are not convinced. Perhaps the most outspoken among this minority group is C. Randy Gallistel:

> We do not yet know in what abstract form (e.g., analog or digital) the mind stores the basic numerical quantities that give substance to the foundational abstractions, the information acquired from experience that specifies learned distances, directions, circadian phases, durations, and probabilities. Much less do we know the physical medium in nervous tissue that is modified in order to preserve these empirical quantities

for use in later computations. (Gallistel, 2016)

**Eric Kandel's Experiment on Sea Slugs: Memories Stored in Synapses**

The groundbreaking work on how memory is (believed to be) stored in the human brain

was performed by the research laboratory of Eric R. Kandel on the sea slug *Aplysia* (Kupfermann

et al., 1970; Pinsker et al., 1970). *Aplysia*, also known as the California brown sea hare, is a

marine snail that has no external shell. It is typically about 20 cm (7.87 inches) in length and 1

kg (2.2 lbs) in weight. Kandel chose *Aplysia* for his study of memory because it has a simpler

model of a nervous system compared to mammals; *Aplysia* only has about 20,000 neurons, in

comparison to about 86 billion neurons in the human brain (Azevedo et al., 2009). *Aplysia's*

simple protective reflex of protecting its gills was instrumental in Kandel's experiment

(Kupfermann et al., 1970; Pinsker et al., 1970). He found that some types of stimuli resulted in

strengthening of the sea slug's protective reflex, signifying learned fear. Strengthening was due

to an amplification of the synapses that connect the sensory neurons to the motor neurons that

produced the protective reflex. Kandel found that weaker stimuli resulted in short-term memory

(shorter duration strengthening of the protective reflex that lasted for minutes to hours), whereas more powerful stimuli resulted in long-term memory (longer duration strengthening that remained for weeks). He also found that long-term memory required new protein to be formed,

whereas short-term memory did not. If the process of synthesizing new protein was blocked, the

formation of long-term memory was also blocked, but not short-term memory. The essence of

Kandel's discovery was that synapses grew/changed when new memories were formed; he consequently interpreted this as evidence that short-term memory and long-term memory in the sea slug were stored in the synapse. During the 1990s, he showed that the same type of long-term growth/changes in the synapses associated with protective reflex (learned fear) in sea slugs also applied to learned fear in the amygdala of mice, and thus, by extension of the animal model, was applicable to humans as well (Kandel, 2006). Kandel concluded that memory in the brain was stored in synapses, and changes to synapses were central to the formation of short-term and long-term memories. In other words, Kandel's experiment provided evidence in support of Hebb's synaptic hypothesis (Hebb, 1949). Based on his discovery of the synapse as the physiological basis of memory storage, Kandel was awarded the 2000 Nobel Prize in Physiology or Medicine (Nobel Prize, 2000).

**Key Experiments on Mice: Declarative Memory and the Hippocampus**

Following Kandel's work in *Aplysia*, the next key experimental findings in support of Hebb's synaptic hypothesis were discovered in mice, the most notable of which was performed by John O'Keefe (O'Keefe, 1976; O'Keefe & Dostrovsky, 1971). Using electrophysiology (the study of electrical properties of biological cells and tissues), he recorded the firing of individual neurons in the hippocampus of mice that were awake and freely moving in a room. O'Keefe discovered that some specific hippocampus neurons were always activated when the mice were at a certain location in the room, whereas other specific hippocampus neurons were activated when the mice were at a different location in the room. Based on this observation, O'Keefe interpreted that the hippocampus contained cognitive maps of the external environment, which the mice utilized for navigation. He named these neurons *place cells*. He concluded that memory of the environment was stored as a combination of these *place cells*. This work subsequently helped define the role of the hippocampus in declarative memory in humans. For example, recent neuroimaging studies have found evidence of the existence of *place cells* in humans as well (Hassabis et al., 2009), especially in patients with Alzheimer's disease where the hippocampus were frequently affected at an early stage which resulted in these patients often losing their way and unable to recognize the environment. For his ground-breaking work on *place cells* in the hippocampus, O'Keefe was awarded the 2014 Nobel Prize in Physiology or Medicine (Nobel Prize, 2014).

The discovery of *place cells* paved the way for the interpretation of results found in a study on rabbits by Bliss and Lomo (1973). By stimulating the hippocampus using a high frequency train of action potentials, they found prolonged/persistent strengthening of the synapses (i.e., long-term potentiation (LTP)) in all three major hippocampal pathways (perforant pathway, mossy fiber pathway, Schaffer collateral pathway). In two of these pathways (perforant and Schaffer collateral), the LTP (strengthening of synapses) was found to be consistent with Hebb's synaptic hypothesis, which consequently reinforced the notion that that memory information was stored in the synapse (Bliss & Collingridge, 1993; Mayford et al., 2012).

**Joseph LeDoux's Experiments on Mice: Procedural Fear Memory and the Amygdala**

Classical conditioning experiments on mice conducted by the research laboratory of Joseph E. LeDoux at NYU found some of the strongest evidence that reinforced Hebb's synaptic hypothesis. Classical conditioning was first studied by Ivan Pavlov, who used his dogs as subjects (Nobel Prize, 1904). Typically, his dogs would salivate when food was presented, but not when a bell was rung. However, if the bell was rung before the food was presented and this sequential process was repeated, his dogs would eventually salivate even when the bell was rung without presentation of the food. Meaning, the bell (conditioned stimulus) resulted in the same response as the food (salivate). In LeDoux's experiments, he paired an audio tone with an electric shock to the feet of mice, which subsequently resulted in a conditioned fear response (freezing behavior) to the audio tone alone (LeDoux, 1995; Rogan et al., 1997). This form of learning (fear conditioning) was known to involve the amygdala, which receives auditory input and regulates autonomic fear responses. LeDoux found that this conditioned fear resulted in LTP (strengthening of synapses) in the auditory neurons of the amygdala, to which he concluded that the LTP constituted memory of the conditioned fear. That is, memory was stored by way of strengthening the synapses, as hypothesized by Hebb.

**Synapse as the Basic Building Block for Artificial Intelligence Neural Networks**

In the adjacent field of artificial intelligence (AI), the concept of synapse serves as the underlying basis for neural networks (NN). At each neuron in the AI NN, there are multiple inputs and multiple outputs; each of the inputs ($x_i$), is weighted (multiplied) by a numerical value ($w_i$), after which all the weighted inputs are summed (added) to produce an output ($y$):

$$y = \sum_i x_i w_i$$

The output ($y$) subsequently serves as the input to other neurons. Here, the synapse (in a brain) is conceptually analogous to the numerical value that each input to the neuron (of the AI NN) is weighted by, also known as the synaptic weight. The sequential cascade (i.e., series interconnection) of one neuron's output serving as another neuron's input in an AI NN is known as a layer. Recent advances in computing power/speed have enabled the use of many such layers, resulting in what is termed Deep Learning (LeCun et al., 2015). AI NN with 20 layers and hundreds of millions synaptic weights have been highly effective in recognizing images and human faces, to the extent that a variant of Deep Learning called Deep Convolutional Neural Networks have been hypothesized to mimic neurons in the visual cortex of the brain (Lindsay, 2020). Deep Learning has also been successfully applied to natural language processing (understanding semantics), most notably in 2011 when IBM's Watson computer defeated two human champions (Ken Jennings, Brad Rutter) in the TV quiz show *Jeopardy!* (Markoff, 2011). In 2017, Google's AlphaGo computer conquered the game Go when it defeated the world's number one Go player, Ke Jie (Mozur, 2017). The Deep Learning approach employed by AlphaGo was a variant known as reinforcement learning (a computer learning method based on the psychological concepts of operant conditioning and reinforcement that were initially proposed by psychologist B. F. Skinner), which has been associated with the dopamine reward system in the brain (Niv, 2009). The impressive feats accomplished by these synapse-based Deep Learning neural networks, along with the hypothesized similarities with the brain (i.e., visual cortex, semantics, dopamine reward system) indirectly supported Hebb's synaptic hypothesis.

**Lingering Doubts on Synapse as the Physical Basis of Memory**

Despite the conceptual similarities, synaptic weights in AI NN have constant values that do not change after the training phase has been completed. In contrast, synapses in the brain are constantly changing, in part due to the inevitable existence of noise (Faisal et al., 2008). Furthermore, AI NN are based on modern computers that function using registers (a type of computer memory used for addition and multiplication mathematical operations), whereas there is no evidence that such registers exist in the brain. Consequently, it would be fair to surmise that the brain is very unlikely to function in the same way as AI neural networks.

For decades, Hebb's synaptic hypothesis, along with key supportive experimental results of the synaptic mechanism of memory, held great promise for the development of new medications/treatments for memory-related illnesses such as Alzheimer's disease (loss of explicit memory). The general idea was that, since memory was stored in the synapse, then, addressing/resolving the synaptic pathology could help treat memory disorders (Jackson, et al.

(2019). However, the long-awaited breakthroughs have yet to be found, raising some doubts against Hebb's synaptic hypothesis and the subsequent associated experimental findings.

One indicative counterevidence arose from the study of motor memory in mice. Using two-photon microscopy, it was found that learning a new motor skill (i.e., new motor memory) was indeed accompanied by the formation of new synaptic connections (Yang et al, 2009). However, unexpectedly, synaptic spines were found to be turning over (changing) at a high rate in the absence of learning, to the extent that newly formed synaptic connections (supposedly encoding new memory) would have vanished in due time, implying that motor memories far outlived their supposed constituent parts (synapses) (Trettenbrein, 2016). This perplexing finding was perhaps best summarized by Emilio Bizzi and Robert Ajemian:

> If we believe that memories are made of patterns of synaptic connections sculpted by experience, and if we know, behaviorally, that motor memories last a lifetime, then how can we explain the fact that individual synaptic spines are constantly turning over and that aggregate synaptic strengths are constantly fluctuating? How can the memories outlast their putative constitutive components? (Bizzi & Ajemian, 2015, pp. 91-92)

They further pointed out that this mystery existed beyond motor neuroscience, extending to all of systems neuroscience given that many studies have found such constant turn over of synapses regardless of the cortical region. In order words, synapses are constantly changing throughout the entire brain: "How is the permanence of memory constructed from the evanescence of synaptic spines?" (Bizzi & Ajemian, 2015, p. 92). This is perhaps the biggest challenge against the notion of synapse as the physical basis of memory.

**David Glanzman's Experiment on Sea Slugs: Memories Not Stored in Synapses**

  Doubts on the synaptic basis for memory were validated in a study conducted by the research laboratory of David L. Glanzman at UCLA, which found that long-term memories could be restored after synapses were pharmacologically eliminated (Chen et al., 2014). It is worth noting that Glanzman was previously a postdoctoral researcher in Eric Kandel's lab at Columbia University. Glanzman grew *Aplysia* neurons in petri dishes and trained/treated them with the hormone serotonin, which subsequently triggered the growth of new synapses as expected and predicted by Kandel's study. After that, the neurons were given pharmacological treatments (anisomycin, chelerythrine) that disrupted long-term memory, and more significantly, reversed the synaptic growth that resulted from the serotonin (the synapses reverted to the way they were before being trained/treated by serotonin). In addition to the reversal, some synapses that existed prior to the serotonin training/treatment were also lost. Based on Hebb's synaptic hypothesis, the long-term memory should have been erased as well, given the reversal of the synaptic growth and loss of synapses. Surprisingly, the long-term memory remained intact. This finding suggested that, while synapses have grown during the formation of long-term memory, storage/recollection of the memory was not dependent on retaining/maintaining the synapses. Thus, these results challenged Hebb's hypothesis that synapses store long-term memories. Glanzman concluded that "long-term memory storage and synaptic change can be dissociated" (Chen et al., 2014, p. 1). For people who suffer from post-traumatic stress disorders (PTSD), this result suggested that the potential use of medications (propranolol) to disrupt the synapses will unlikely eliminate painful memories. At the same time, this result offered some hope to people who suffer from dementia or Alzheimer's disease; some part of the memories may be recoverable despite the neurodegeneration (deterioration/loss of synapses).

**Susumu Tonegawa's Experiment on Mice: Reactivating Retrograde Amnesia Using Laser**

Further evidence against Hebb's synaptic hypothesis was reported by Susumu Tonegawa of MIT. In an experiment conducted by Tonegawa's research lab (Ryan et al., 2015), neurons in conditioned/trained mice were injected with a pharmacological treatment (anisomycin), which disrupted the synaptic growth/consolidation (that Kandel deemed necessary for memory storage). Consequently, retrograde amnesia was induced (the memory could not be retrieved by the mice via an emotional/fear trigger). However, these "lost" memories could be reactivated by shining laser onto the corresponding (memory) neurons that were tagged during the conditioning/training stage. Here, laser refers to optogenetics, a biological technique that employs light to control neurons that have been genetically modified to express light-sensitive ion channels. Tonegawa's experiment on mice was, in essence, a replication of Glanzman's experiment on sea slugs; in both cases, the animals were trained/conditioned, and then, pharmacological treatments (anisomycin) were used to disrupt the growth of synapses, which, according to Hebb's synaptic hypothesis, should have erased the memory; but in both cases, memory remained retrievable despite the pharmacological blocking of the synapse. Tonegawa's study concluded that an increase in synaptic strength was not a crucial requisite for storage of memory information. This further reinforced the doubts on Hebb's synaptic hypothesis cast by Bizzi and Ajemian (2015).

**Germund Hesslow's Experiments on Ferrets: LTP Cannot Explain Conditioned Pauses**

Pavlovian eye-blink conditioning experiments on ferrets conducted by the research laboratory of Germund Hesslow at Lund University raised further doubts on Hebb's synaptic hypothesis (Johansson et al., 2014). Typically, the eye would blink in response to presentation of an air puff, similar to the way Pavlov's dog would salivate in response to presentation of food. In Hesslow's study, the air puff was paired with an electrical pulse to the paw of the ferret, analogous to the bell in Pavlov's study. Prior to conditioning, the electrical pulse to the paw produced no eye blinks; after conditioning (stimulating the paw with an electrical pulse before presenting the air puff), the electrical pulse to the paw produced eye blinks even in the absence of an air puff. Hesslow measured the electrophysiological responses of Purkinje cells in the cerebellum that were associated with eye-blinks in order to examine how the cells would respond to the paired stimulus (electrical pulse to the paw). Prior to conditioning, the electrical pulse to the paw did not change the firing pattern of the Purkinje neurons. After conditioning, a 200-millisecond electrical pulse to the paw resulted in an approximately 200-millisecond pause in the Purkinje cells' neural spike activities; likewise, a 300-millisecond electrical pulse resulted in an approximately 300-millisecond pause. These findings indicated that the Purkinje cell neurons were able to remember the time duration (e.g., 200-millisecond, 300-millisecond) of the paired stimulus (electrical pulse to the paw) in a rather precise and proportionate manner. Hesslow concluded that LTP (strengthening of synapses) could not account for the Purkinje cells' ability to remember the time durations: "Mere strengthening or weakening of these synapses cannot account for the time course of the conditioned pause response" (Johansson et al., 2014, pp. 14932). Consequently, Hesslow's experiments further added doubts to Hebb's synaptic hypothesis.

**Beatrice Gelber's Experiments on Paramecium: Conditioning Without Synapses**

Pavlovian conditioning experiments on *Paramecium aurelia* in the 1950s, conducted by psychologist Beatrice Gelber at Indiana University and the University of Chicago, raised further questions on Hebb's synaptic hypothesis (Gelber, 1957). *Paramecium aurelia* is a single-cell organism, typically oblong or slipper-shaped, covered in hair-like filaments (cilia). Most people would remember encountering *Paramecia* at some point in high school science classes, by way of peering through a microscope. Gelber was interested in finding out whether simple single-cell organisms such as *Paramecia* were capable of Pavlovian conditioning, which was (and still is) widely considered a sophisticated form of learning. One of her astonishing findings ended up being published in *Science* (Gelber, 1957). In that study, a microdrop of bacterial suspension (i.e., food) was introduced at the edge of a container which had a "hungry" culture of *Paramecia*. In the experimental group, a clean wire was simultaneously lowered into the middle of the container; after 8 minutes, the wire was removed. The control group received the food without the wire. After 30 minutes, a clean and sterile wire was introduced in each of the cultures/containers. Gelber found that *Paramecia* in the experimental group surrounded the wire significantly more than those in the control group ($p < .02$). Based on this result, along with other variations of experimental design, she concluded that *Paramecium aurelia* was indeed capable of Pavlovian conditioning. Despite the gravitas of this discovery, Gelber's studies were ignored and/or dismissed by her contemporaries, and largely forgotten until earlier this year (January 2021) when Harvard psychologist Samuel J. Gershman brought Gelber's work back into the spotlight (Gershman et al., 2021). Barring Hesslow's studies on ferrets (Johansson et al., 2014), the prevailing theory is that Pavlovian conditioning is mediated by Hebb's synaptic hypothesis. However, single cell organisms clearly do not have synapses; if *Paramecia* can be conditioned to remember, they must be using a non-synaptic form of memory storage. Therefore, synapses may not actually be essential for memory storage, calling Hebb's synaptic hypothesis into question.

**Alternatives to Hebb's Synaptic Hypothesis**

The logical question to pose at this point is: if memory information is not stored in the synapse, then where is it? Glanzman suggested that memory might be stored in the nucleus of the neurons (Chen et al., 2014). On the other hand, Tonegawa proposed that memory might be stored in the connectivity pathways (circuit connections) of a network of neurons (Ryan et al., 2015).

Hesslow emphasized that memory is highly unlikely to be a network property (in disagreement

with Tonegawa), and further posited that the memory mechanism is intrinsic to the neuron (in

agreement with Glanzman) (Johansson et al., 2014). Decades earlier, Gelber (1962) hypothesized

that memory is "coded in macromolecules" (p. 166) (inside the cell of the *Paramecia*), and she further postulated that "the biochemical and cellular physiological processes which encode new responses are continuous across the phyla" (p. 166), implying that the memory mechanisms would be "reasonably similar for a protozoan and a mammal" (p. 166). Gershman expressed a cautious agreement with Gelber that, "if the hypothesis is correct, then single cells hold more surprises in store for us" (Gershman et al., 2021, p. 11). The collective views of Glanzman,

Hesslow, Gelber, and Gershman is known as the *cell-intrinsic hypothesis* – that, memory information is stored in information-bearing molecules inside the neuron (Gallistel, 2017).

**Plausibility of the Cell-Intrinsic Hypothesis**

Peter Sterling (University of Pennsylvania) and Simon Laughlin (University of

Cambridge) suggested that storing memory and performing computations using molecular

chemistry inside the neuron would be energetically cheaper in comparison to using neural spikes

and synapses (Hebb's synaptic hypothesis) (Sterling & Laughlin, 2015). Gershman further

elaborated that "a synaptic memory substrate requires that computations operate via the

propagation of spiking activity, incurring an energetic cost roughly 13 orders of magnitude

greater than the cost incurred if the computations are implemented using intracellular molecules"

(Gershman et al., 2021, p. 2). It is worth noting here that 13 orders of magnitude equate to $10^{13}$,

suggesting that synaptic memory would require approximately 10 trillion times more energy than molecular memory. Within the neuron, there are two major types of molecules that are known to be capable of storing information: deoxyribonucleic acid (DNA), and ribonucleic acid (RNA) (Gallistel, 2017).

Francis Crick, who was awarded the 1962 Nobel Prize in Physiology or Medicine for deciphering the helical structure of the DNA molecule (Nobel Prize, 1962), was first to suggest that "memory might be coded in alternations to particular stretches of chromosomal DNA" (Crick, 1984, p. 101). The hypothesized epigenetic (non-genetic influences on gene expression) mechanism for memory (DNA methylation or demethylation) was further elaborated by molecular biologist Robin Holliday (Holliday, 1999). Recent work by researchers at Johns Hopkins University School of Medicine (Yu et al., 2015) concluded that neurons constantly rewrite their DNA: "We used to think that once a cell reaches full maturation, its DNA is totally stable" but "this research shows that some cells actually alter their DNA all the time, just to perform everyday functions" (Johns Hopkins Medicine, 2015). In a collaborative effort among researchers at the University of Alabama at Birmingham, Bates College, and Vanderbilt University, 9.2% of DNA in the hippocampus of mice were found to be altered after fear conditioning (Duke et al, 2017). Another recent work (McConnell et al., 2017) concluded that no two neurons are genetically alike: "We were taught that every cell has the same DNA, but that's not true" because "neural genes are very active" (Makin, 2017).

All single-stranded RNA in the cell is made from double-stranded DNA, via a process called transcription (Alberts et al., 2002). Consequently, changes in the DNA would be passed on to the RNA. Alternatively, RNA could also potentially be altered on its own, without necessarily involving the DNA. It is worth noting here that there are many types of RNA (messenger RNA, transfer RNA, ribosomal RNA, microRNA). An RNA-based hypothesis of memory and computation has recently been proposed by Hessameddin Akhlaghpour of The Rockefeller University (Akhlaghpour, 2020).

**David Glanzman's Experiment: Transplanting Memory Between Sea Slugs Using RNA**

Glanzman conducted a follow up experiment (Bedecarrats et al., 2018) to test the cell-intrinsic hypothesis – specifically, on memory information storage in RNA molecules inside the neuron. *Aplysia* sea slugs were given repeated mild electric shocks to their tails (experimental group), resulting in an enhanced defensive withdrawal reflex to protect from potential harm. Subsequently, when those sea slugs were tapped, their defensive withdrawal response averaged 56 seconds in duration. On the other hand, sea slugs that did not previously receive electric shocks (control group) responded for only about 1 second. RNA from both groups were subsequently extracted. RNA from the experimental group was injected into one new group of naïve sea slugs (sea slugs that have never received any electric shock), whereas RNA from the control group was injected into another new group of naïve sea slugs. Glanzman found that the group of naïve sea slugs that received RNA from the control group exhibited a defensive withdrawal response of about 5 seconds. Remarkably, the group of naïve sea slugs that received RNA from the experimental group had a response of about 38 seconds. In other words, naïve sea slugs that received RNA from the experimental group responded as if they themselves had received electric shocks, displaying a response duration that was similar in length to those that actually received electric shocks (experimental group). Glanzman attributed the longer response duration to the RNA injection, and concluded that "it's as though we transferred the memory" because "if memories were stored at the synapses, there is no way our experiment would have worked" (University of California, Los Angeles, 2018).

**Dias and Ressler's Experiment on Mice: DNA Transfer of Fear from Parents to Offspring**

An experiment conducted by Brian G. Dias and Kerry J. Ressler at Emory University found that fear conditioning in mice could be transferred from parents to offspring (Dias & Ressler, 2014). Using Pavlovian conditioning, they trained mice to be fearful of a scent (acetophenone, which smelled like cherry blossom) by pairing it with a mild electric shock. After conditioning, the mice learnt to associate the scent with pain, startling in the presence of the scent even without an electric shock. They found that offspring of the conditioned mice startled more in response to the scent, even though the offspring were naïve (not previously conditioned to associate the scent with pain from an electric shock). Astonishingly, the sensitivity was also observed in the second-generation mice (grandchildren). Dias and Ressler concluded that the conditioned fear associated with the scent was transferred to the offspring via DNA in the sperm or eggs of the mice, suggesting that the offspring inherited the fear from their parents. In short, traumatic memories could be inherited, at least in mice. Ressler suggested that humans may also inherit epigenetic alterations that influence behavior: "A parent's anxiety could influence later generations through epigenetic modifications to receptors for stress hormones" (Callaway, 2013). He added that "knowing how the experiences of parents influence their descendants helps us to understand psychiatric disorders that may have a transgenerational basis, and possibly to design therapeutic strategies" (Eastman, 2013).

## Conclusions

After more than 70 years of research efforts by cognitive psychologists and neuroscientists, the question of where memory information is stored in the brain remains unresolved. Although the long-held synaptic hypothesis remains as the de facto and most widely accepted dogma, there is growing evidence in support of the cell-intrinsic hypothesis. As Glanzman summed up rather succinctly, "I expect a lot of astonishment and skepticism" (McFarling, 2018). In a recent interview in April 2021, Gallistel was quoted saying "Scientists are human. Like all humans, they're prisoners of preconceptions. When a preconception takes strong hold, it becomes almost unshakable" (Join Activism, 2021). He further reiterated a famous quote by physicist Max Planck that "science progresses one funeral at a time" (Join Activism, 2021).

A synapse connects one neuron to another. Without synapses, (most) neurons would not be able to communicate with one another; sensory information (e.g., from the retina) would not reach the brain in the first place. Consequently, both the synapse and the cell are likely to be crucial to memory, with each serving a potentially different but inter-dependent function: while the cell might be storing the memory information, the synapse might be required for the initial formation and the subsequent retrieval of the memory (Tee & Taylor, 2021). A (potentially) helpful analogy here is the way a road leads to a warehouse that stores goods; while the warehouse stores the goods, the road is required for the initial delivery and subsequent pickup of the goods. Following this analogy, it would be risky to store all goods in just one warehouse (in case of fire or burglary). Furthermore, there is a finite amount of storage space/capacity in each warehouse. Therefore, it would be wise and/or inevitable to store goods across multiple warehouses that are interconnected by a network of roads. When goods are picked up from the multitude of warehouses, the complex logistical process may not always result in a perfect retrieval of the expected quantity or type of goods. Likewise in the brain, it would make sense to store memory information across multiple neurons that are interconnected by a network of synapses. When memories (e.g., episodic, autobiographical) are retrieved from the multitude of neurons, the complex recollection process may not always result in a perfect retrieval. Such a model could potentially account for errors of omission (forgetting information) and errors of commission (remembering the wrong information) in reconstructive memory.

      Lastly, if DNA is indeed involved in the storage of long-term memory (in humans), there

are profound implications beyond neuroscience and cognitive psychology. For example, could memories (PTSD, substance use, racial discrimination) be inherited from one generation to another? If so, how would such inherited memories affect members of a community (collective memory)? These types of open research questions have far-reaching ramifications for clinical, developmental, and social psychology.

**References**

Akhlaghpour, H. (2020). A theory of universal computation through RNA. *arXiv.*

   https://arxiv.org/abs/2008.08814

Alberts, B., Johnson, A. D., Lewis, J., Raff, M., Roberts, K., & Walter, P. (2002). From DNA to

   RNA. *Molecular Biology of the Cell* (4th ed.). Garland Science.

   https://www.ncbi.nlm.nih.gov/books/NBK26887/

Azevedo, F. A. C., Carvalho, L. R. B., Grinberg, L. T., Farfel, J. M., Ferretti, R. E. L., Leite, R. E.

   P., Filho, W. J., Lent, R., & Herculano-Houzel, S. (2009). Equal numbers of neuronal and

   nonneuronal cells make the human brain an isometrically scaled-up primate brain. *The Journal*

   *of Comparative Neurology, 513*(5), 532-541. https://doi.org/10.1002/cne.21974

Bedecarrats, A., Chen, S., Pearce, K., Cai, D., & Glanzman, D. L. (2018). RNA from trained

   *Aplysia* can induce an epigenetic engram for long-term sensitization in untrained *Aplysia.*

   *eNeuro, 5*(3), 1-11. https://doi.org/10.1523/ENEURO.0038-18.2018

Bizzi, E., & Ajemian, R. (2015). A hard scientific quest: understanding voluntary movements.

   *Daedalus, 144*(1), 83-95. https://doi.org/10.1162/DAED_a_00324

Bliss, T. V. P., & Lomo, T. (1973). Long-lasting potentiation of synaptic transmission in the

   dentate area of the anaesthetized rabbit following stimulation of the perforant path. *The*

   *Journal of Physiology, 232*(2), 331-356. https://doi.org/10.1113/jphysiol.1973.sp010273

Bliss, T. V. P., & Collingridge, G. L. (1993). A synaptic model of memory: Long-term

   potentiation in the hippocampus. *Nature*, *361*(6407), 31-39. https://doi.org/10.1038/361031a0

Callaway, E. (2013, December 1). Fearful memories haunt mouse descendants. *Nature News.*

   https://www.nature.com/news/fearful-memories-haunt-mouse-descendants-1.14272

Chen, S., Cai, D., Pearce, K., Sun, P. Y.-W., Roberts, A. C., & Glanzman, D. L. (2014).

Reinstatement of long-term memory following erasure of its behavioral and synaptic

expression in Aplysia. *eLife, 3*(e03896), 1-21. https://doi.org/10.7554/eLife.03896

Crick, F. (1984). Memory and molecular turnover. *Nature, 312*(5990), 101.

https://doi.org/10.1038/312101a0

Dias, B. G., & Ressler, K. J. (2014). Parental olfactory experience influences behavior and neural structure in subsequent generations. *Nature Neuroscience, 17*(1), 89-96.

https://doi.org/10.1038/nn.3594

Duke, C. G., Kennedy, A. J., Gavin, C. F., Day, J. J., & Sweatt, J. D. (2017). Experience-

dependent epigenomic reorganization in the hippocampus. *Learning & Memory, 24*(7), 278-

288. https://doi.org/10.1101/lm.045112.117

Eastman, Q. (2013, December 2). Mice can inherit learned sensitivity to a smell. *Emory News

Center*. https://news.emory.edu/stories/2013/12/smell_epigenetics_ressler/campus.html

Faisal, A. A., Selen, L. P. J., & Wolpert, D. M. (2008). Noise in the nervous system. *Nature

Reviews Neuroscience, 9*(4), 292-303. https://doi.org/10.1038/nrn2258

Gallistel, C. R. (2016). Presidential column: What we have and haven't learned. *Observer, 29*(5).

https://www.psychologicalscience.org/observer/what-we-have-and-havent-

learned#.WSFTuhPyuRs

Gallistel, C. R. (2017). The coding question. *Trends in Cognitive Sciences, 21*(7), 498-508.

https://doi.org/10.1016/j.tics.2017.04.012

Gelber B. (1957). Food or training in Paramecium. *Science, 126*(3287)*,* 1340-1341.

https://doi.org/10.1126/science.126.3287.1340

Gelber, B. (1962). Acquisition in Paramecium aurelia during spaced training. *The Psychological Record, 12*(2), 165-177. https://doi.org/10.1007/BF03393454

Gershman, S. J., Balbi, S. E. M., Gallistel, C. R., & Gunawardena, J. (2021). Reconsidering the evidence for learning in single cells. *eLife, 10*(e61907), 1-15. https://doi.org/10.7554/eLife.61907

Hassabis, D., Chu, C., Rees, G., Weiskopf, N., Molyneux, P. D., & Maguire, E. A. (2009). Decoding neuronal ensembles in the human hippocampus. *Current Biology, 19*(7), 546-554. https://doi.org/10.1016/j.cub.2009.02.033

Hebb, D. O. (1949). *The Organization of Behavior: A Neuropsychological Theory*. Wiley and Sons.

Holliday, R. (1999). Is there an epigenetic component in long-term memory? *Journal of Theoretical Biology, 200*(3), 339-341. https://doi.org/10.1006/jtbi.1999.0995

Jackson, J., Jambrina, E., Li, J., Marston, H., Menzies, F., Phillips, K., & Gilmour, G. (2019). Targeting the synapse in Alzheimer's disease. *Frontiers in Neuroscience, 13*, 735. https://doi.org/10.3389/fnins.2019.00735

Johansson, F., Jirenhed, D.-A., Rasmussen, A., Zucca, R., & Hesslow, G. (2014). Memory trace and timing mechanism localized to cerebellar Purkinje cells. *Proceedings of the National Academy of Sciences of the United States of America, 111*(41), 14930-14934. https://doi.org/10.1073/pnas.1415371111

Johns Hopkins Medicine (2015, April 27). *Neurons constantly rewrite their DNA*. https://www.hopkinsmedicine.org/news/media/releases/neurons_constantly_rewrite_their_dna

Join Activism (2021, April 12). *Is this the most interesting idea in all of science?* https://join.substack.com/p/is-this-the-most-interesting-idea

Kandel, E. R. (2006). *In search of memory: The emergence of a new science of mind.* W. W. Norton & Co.

Kupfermann, I., Castellucci, V., Pinsker, H., & Kandel, E. (1970). Neuronal correlates of habituation and dishabituation of the gill-withdrawal reflex in *Aplysia*. *Science, 167*(3926), 1743-1745 (1970). http://doi.org/10.1126/science.167.3926.1743

Lashley, K. S. (1950). In search of the engram. In Society for Experimental Biology, *Symposium IV: Physiological mechanisms in animal behavior* (pp. 454-482). Academic Press.

LeCun, Y., Bengio, Y., & Hinton, G. (2015). Deep learning. *Nature, 521*(7553), 436-444. https://doi.org/10.1038/nature14539

LeDoux, J. E. (1995). Emotion: Clues from the brain. *Annual Review of Psychology, 46*, 209-235. https://doi.org/10.1146/annurev.ps.46.020195.001233

Lindsay, G. W. (2020). Convolutional neural networks as a model of the visual system: Past, present, and future. *Journal of Cognitive Neuroscience,* advance online publication, 1-15. https://doi.org/10.1162/jocn_a_01544

Makin, S. (2017, May 3). Scientists surprised to find no two neurons are genetically alike. *Scientific American.* https://www.scientificamerican.com/article/scientists-surprised-to-find-no-two-neurons-are-genetically-alike/

Markoff, J. (2011, February 16). Computer wins on 'Jeopardy!': Trivial, it's not. *The New York Times.* https://www.nytimes.com/2011/02/17/science/17jeopardy-watson.html

Mayford, M., Siegelbaum, S. A., & Kandel, E. R. (2012). Synapses and memory storage. *Cold Spring Harbor Perspectives in Biology, 4*(a005751), 1-18. https://doi.org/10.1101/cshperspect.a005751

McConnell, M. J., Moran, J. V., Abyzov, A., Akbarian, S., Bae, T., Cortes-Ciriano, I., Erwin, J.
A., Fasching, L., Flasch, D. A., Freed, D., Ganz, J., Jaffe, A. E., Kwan, K. Y., Kwon, M.,
Lodato, M. A., Mills, R. E., Paquola, A. C. M., Rodin, R. E. Rosenbluh, C., …Brain Somatic
Mosaicism Network (2017). Intersection of diverse neuronal genomes and neuropsychiatric
disease: The Brain Somatic Mosaicism Network. *Science, 356*(6336), 1-9.
https://doi.org/10.1126/science.aal1641

McFarling, U. L. (2018, May 14). Memory transferred between snails, challenging standard
theory of how the brain remembers. *Scientific American.*
https://www.scientificamerican.com/article/memory-transferred-between-snails-challenging-
standard-theory-of-how-the-brain-remembers

Mozur, P. (2017). Google's AlphaGo defeats Chinese Go Master in Win for A.I. *The New York
Times. https://www.nytimes.com/2017/05/23/business/google-deepmind-alphago-go-
champion-defeat.html*

Niv, Y. (2009). Reinforcement learning in the brain. *Journal of Mathematical Psychology, 53*(3),
139-154. https://doi.org/10.1016/j.jmp.2008.12.005

Nobel Prize (1904). *The Nobel Prize in Physiology or Medicine 1904: Nobel Lecture by Ivan
Pavlov.* https://www.nobelprize.org/prizes/medicine/1904/pavlov/lecture/

Nobel Prize (1962). *The Nobel Prize in Physiology or Medicine 1962.*
https://www.nobelprize.org/prizes/medicine/1962/summary/

Nobel Prize (2000). *The Nobel Prize in Physiology or Medicine 2000.*
https://www.nobelprize.org/prizes/medicine/2000/press-release/

Nobel Prize (2014). *The Noble Prize in Physiology or Medicine 2014.*
https://www.nobelprize.org/prizes/medicine/2014/press-release/

O'Keefe, J. (1976). Place units in the hippocampus of the freely moving rat. *Experimental Neurology, 51*(1), 78-109. https://doi.org/10.1016/0014-4886(76)90055-8

O'Keefe, J., & Dostrovsky, J. (1971). The hippocampus as a spatial map. Preliminary evidence from unit activity in the freely moving rat. *Brain Research*, *34*(1), 171-175. https://doi.org/10.1016/0006-8993(71)90358-1

Pinsker, H., Kupfermann, I., Castellucci, V. & Kandel, E. (1970). Habituation and dishabituation of the gill-withdrawal reflex in Aplysia. *Science, 167*(3926), 1740-1742. https://doi.org/10.1126/science.167.3926.1740

Rogan, M. T., Staubli, U. V., & LeDoux, J. E. (1997). Fear conditioning induces associative long-term potentiation in the amygdala. *Nature, 390*(6660), 604–607. https://doi.org/10.1038/37601

Ryan, T. J., Roy, D. S., Pignatelli, M., Arons, A., & Tonegawa, S. (2015). Engram cells retain memory under retrograde amnesia. *Science, 348*(6238), 1007-1013. https://doi.org/10.1126/science.aaa5542

Shatz, C. J. (1992). The developing brain. *Scientific American, 267*, 60–67. https://doi.org/10.1038/scientificamerican0992-60

Sterling, P., & Laughlin, S. B. (2015). *Principles of Neural Design*. MIT Press.

Tee, J., & Taylor, D. P. (2021). What if memory information is stored inside the neuron, instead of in the synapse? *arXiv*. https://arxiv.org/abs/2101.09774

Trettenbrein P. C. (2016). The demise of the synapse as the locus of memory: A looming paradigm shift? *Frontiers in Systems Neuroscience, 10*(88), 1-7. https://doi.org/10.3389/fnsys.2016.00088

University of California, Los Angeles (2018, May 14). *UCLA biologists "transfer" a memory*.

https://newsroom.ucla.edu/releases/ucla-biologists-transfer-a-memory

Yang, G., Pan, F., & Gan, W.-B. (2009). Stably maintained dendritic spines are associated with

lifelong memories. *Nature, 462*(7275), 920-924. https://doi.org/10.1038/nature08577

Yu, H., Su, Y., Shin, J., Zhong, C., Guo, J. U., Weng, Y.-L., Gao, F., Geschwind, D. H.,

Coppola, G., Ming, G.-L., & Song, H. (2015). Tet3 regulates synaptic transmission and homeostatic plasticity via DNA oxidation and repair. *Nature Neuroscience, 18*(6), 836-845.

https://doi.org/10.1038/nn.4008